\documentclass[preprint,epsf]{revtex4}
\usepackage[dvips]{graphicx}

\begin{document}

\title{Inferring genetic networks: an information theoretic
approach}

\author{L. Diambra\footnote{E-mail: ldiambra@creg.org.ar}}
\address{Laboratorio de Biolog\'{\i}a de Sistemas -- CREG-UNLP, \\
Av. Calchaqui Km 23.5 CP 1888, Florencio Varela, Argentina.}

\begin{abstract}
In the postgenome era many efforts have been dedicated to
systematically elucidate the complex web of interacting genes and
proteins. These efforts include experimental and computational
methods. Microarray technology offers an opportunity for monitoring
gene expression level at the genome scale. By recourse to
information theory, this study proposes a mathematical approach to
reconstruct gene regulatory networks at coarse-grain level from high
throughput gene expression data. The method provides the {\it a
posteriori} probability that a given gene regulates positively,
negatively or does not regulate each one of the network genes. This
approach also allows the introduction of prior knowledge and the
quantification of the information gain from experimental data used
in the inference procedure. This information gain can be used to
chose genes to be perturbed in subsequent experiments in order to
refine the knowledge about the architecture of an underlying gene
regulatory network. The performance of the proposed approach has
been studied by {\it in numero} experiments. Our results
suggest that the approach is suitable for focusing on size-limited
problems, such as, recovering a small subnetwork of interest by
performing perturbation over selected genes.
\end{abstract}

%keywords: gene network, reverse engineering, information theory 
\maketitle

%\section*{Author Summary}
%An important open question of systems biology is how genetic and
%molecular components interact to create different phenotypes. Since
%the advent of DNA microarray technology, various efforts have been
%dedicated to inferring effective gene regulatory network interactions
%from array data. However, the possibility of obtaining a phenomenological
%model of a genome-scale interaction network appears remote, due
%to in all cases of data-driven parameter estimation, the number
%of model parameters estimable from a set of data is strictly limited
%by the size of that set and the level of noise. One way around this
%problem is to reduce the dimensionality of search space. This paper
%presents an approach that allow: i) to integrate into the inference
%procedure genomic data based on experimental genetics, ii) to asses
%the likelihood of the inferred paths, and iii) quantify the information
%carried by the experimental data. Additionally, it is illustrated how
%the method can be used to suggest a series of gene perturbation
%experiments in order to identify pattern of specific subnetworks.

\newpage

\section*{Introduction}
Gene expression is regulated by proteins that enhance or
block polymerase binding at the promoter region. These biochemical
reactions constitute the edges of the gene regulatory networks. One
of the key issues in modern biology is the elucidation of the
structure and function of gene regulatory circuits at the system
level \cite{hood}. To address this challenge many efforts have been
devoted to the task of developing computational methods capable of
inferring the interaction between genes from expression levels both
on small pathways \cite{luna,jjc} as on genome-wide scale (see
\cite{rev1} for a review). Several models for gene regulatory
networks have been proposed in order to infer network interactions
\cite{holter,tegner}, such as Bayesian networks
\cite{friedman,husmeier,beal}, Boolean networks \cite{akutsu},
linear model \cite{dhaeseleer1,dhaeseleer2,vansomeren}. Once a
regulatory network model has been chose, it is possible, in
principle, to recover its parameters with some accuracy. Of course,
more detailed models will require more extensive experimental data.
In general this data is not available for the genome-wide scale
assuming complex model. However, we can concentrate on simpler task,
such as: who is regulating whom? and, Is that an up-regulation or a
down-regulation? The idea behind restricting our questions at this
qualitative information level, is to reduce the amount of data
needed to infer valuable and robust biological knowledge even when
dealing with noise data. In any case, the detailed information
offered by more detailed modeling is not useful without a careful
significance analysis of these predictions. In this sense, this
study proposes a mathematical approach to infer gene networks at the
coarse grain level. The inference process is to be accomplished
according to Ockham's razor, i.e., with the minimum number of
assumptions compatible with the available data. To do that, the
information theory (IT) is used within the framework of the maximum
entropy principle \cite{sh,j}. IT has proved to be of utility in
devising techniques for analyzing gene expression and network
reconstruction \cite{lezon,martinez}, where gene expression levels
were regarded as random variables. Here, complementing these
previous works, each putative interaction has been considered as a
random variable. {\it In numero} experiments show that, in this
case, the IT parlance also provides a powerful framework to discuss
questions related to the modeling process such as: (i) how to
incorporate {\it a priori} information about the gene interaction;
(ii) how to asses the likelihood of the inferred paths; (iii) how to
quantify the information provided by the experimental data; and  (iv)
how to design experiments in order to identify subnetworks.

\section*{The IT approach}

In general a genetic network can be modeled by a set of non-linear
differential equations $\dot{x}_i =f_i\left( x_1(t), \ldots,
x_N(t)\right)$, where $x_i(t)$ is the expression level of gene $i$
at time $t$ and $f_i$ is the regulatory function governing the
expression of gene $i$ \cite{yang}. Near a steady state the
nonlinear system can be approximated by a set of linear differential
equations, $\dot{{\bf x}} = {\bf W} {\bf x}$ where ${\bf W}$, is a
weighted connectivity matrix \cite{yeung}.
In order to uncover the  connectivity matrix, we can apply a
stimulus ${\bf b}=\left(b_1(t), \ldots, b_N(t)\right)^T$, then
measure simultaneously the mRNA levels relative to $N$ genes ${\bf
x}$ and estimate the derivative $\dot{\bf x}$. Repeating the
procedure $M$ times we get a measurement matrix ${\bf X}$ where
columns denote the experiments and where rows indicate individual
genes. Thus, we can approximate the dynamics by
\begin{equation}
\dot{{\bf X}}={\bf W} \  {\bf X} +{\bf B} \label{eje}
\end{equation}
where $\dot{ \bf X}$ and ${\bf B}$ follow the same notation as ${\bf
X}$.

Usually, inferring genetic network attempts to retrieve the weight
matrix ${\bf W}$, where the elements $w_{ij}$ describe the type and
strength of influence of gene $j$ on gene $i$ ($w_{ij}>0$ indicates
activation, $w_{ij}>0$ indicates repression, and a zero indicates no
influence). However, without a careful assessment of the
significance of the weight-values, this could lead to the conclusion
that the network is fully connected in contradiction to the
well-known fact that gene regulatory networks are sparse networks.

In the present work, the maximum entropy principle is applied to
obtain the probability distribution from the data $D_M=\left\{ {\bf
X,\dot{X},B } \right\}$,
 over the possible matrix ${\bf W}$. After
that, using maximum likelihood criterion, the gene interaction
matrix ${\bf I}$ is selected. The elements $I_{ij}$ can take only
three values, depending on the type of influence of gene $j$ on gene
$i$, $I_{ij}=1$ for activation (direct or indirect), $I_{ij}=-1$ for
repression and $I_{ij}=0$ when gene $j$ does not have influence on
gene $i$. In order to infer weights consistent with $D_M$, it is
assumed that each set of weights ${\bf W}$ is realized with
probability $P\left( {\bf W}|D_M\right) $. In other words, a
normalized probability distribution is introduced over the possible
sets ${\bf W}$, which satisfy
\begin{equation} \label{vm}
\langle {\bf W} \rangle =\int P\left( {\bf W}|D_M\right) {\bf W}
d{\bf W}.
\end{equation}
The relative entropy related to an {\em a priori} probability
distribution $P_0$, is given by
\begin{equation}  \label{ent}
H_r\left( D_M\right|P_0)=-\int P\left({\bf W}|D_M\right)\ln \left[
\frac{P\left({\bf W}|D_M\right)}{P_0({\bf W})}\right] d{\bf W},
\end{equation}
where $P_0\left( {\bf W}\right) $ is an appropriate {\em a priori}
distribution. The negative relative entropy $H_r$, known as
Kullback-Leibler distance \cite{lt}, defines the information gained
after $D_M$ has been used in the inference procedure. Thus, in this
framework, the inference process takes place through a modification
of the probability distribution on weights space due to incoming
data.

Among all possible distributions $P\left( {\bf W}|D_M\right) $
consistent with $D_M$, $P\left( {\bf W}|D_M\right)$ has been
selected which comprises no unjustified prejudice. Thus, following
the central tenets of the maximum entropy principle, relative
entropy is maximized subject to the constraints Eq. \ref{vm}.
Thus, the {\it a posteriori} probability distribution yields,
\begin{equation}
P\left( {\bf W}|D_M\right) =\exp \left( -\left( 1+\lambda _0\right)
\right) \exp \left( -{\bf W}\cdot {\bf \Gamma } \right) P_0\left(
{\bf W} \right) , \label{dos}
\end{equation}
where $\lambda _0$ is Lagrange multiplier associated to the
normalization condition, and ${\bf \Gamma} $ the Lagrange
multipliers associated to the constraints Eq. \ref{vm}, which are
determined once $P_0$ is properly selected.

In order to select $P_0$, it is assumed that the weights are
restricted to the values of $I_{ij}$ i.e.  $w_{ij}=0,\pm 1$ and then
a three-peaked {\it a priori} distribution is used, which is
described by
%\begin{equation}
\begin{eqnarray}
P_0\left( {\bf W} \right)= &&(2\pi a)^{-N/2} \prod _{ij}^N \left[
p_{ij}^{0}e^{-\frac{w_{ij}^2}{2a}}
+ p_{ij}^{+}e^{ -\frac{\left(w_{ij}-1 \right)^2}{2a}} + \right. \nonumber  \\
&+& \left. p_{ij}^{-}e^{-\frac{\left(w_{ij}+1 \right)^2}{2a}} \right],
\label{bin}
\end{eqnarray}
%\end{equation}
where $p_{i,j}^{\rm x}$ is the {\it a priori} probability for gene
$j$ to regulate positively (${\rm x}=+$), negatively (${\rm x}=-$)
or to not regulate (${\rm x}=0$) gene $i$. Of course $ p_{ij}^0 +
p_{ij}^+ + p_{ij}^-=1$ for each pair $i,j$. The parameter $a$ can be
regarded as a constraint smoothness parameter. Replacing this choice
in Eq. \ref{dos} the {\it a posteriori} probability distribution
is obtained as a sum of three Gaussians,
%\begin{eqnarray}
%\label{binp} P\left( {\bf W}|D_M \right)&=&\frac{1}{(2\pi
%a)^{N/2}}\prod _{ij}^N z_{ij}^{-1} \left[ p_{ij}^{0}e^{
%-\frac{\left(w_{ij}+a\Gamma _{ij} \right)^2}{2a}} + \right.
%\nonumber \\ && \left. + p_{ij}^{+} e^{ -\Gamma _{ij}}
%e^{-\frac{\left(w_{ij}+a\Gamma _{ij}-1 \right)^2}{2a}} + \right.
%\nonumber \\ && \left. + p_{ij}^{-} e^{\Gamma _{ij}} e^{
%-\frac{\left(w_{ij}+a\Gamma _{ij}+1 \right)^2}{2a}} \right]
%\end{eqnarray}
%\begin{eqnarray}
%\label{binp} P\left( {\bf W}|D_M \right)&=&\frac{1}{(2\pi
%a)^{N/2}}\prod _{ij}^N \left[ p_0(i,j) e^{
%-\frac{\left(w_{ij}+a\Gamma _{ij} \right)^2}{2a}} + \right.
%\nonumber \\ && \left. + p_+(i,j) e^{-\frac{\left(w_{ij}+a\Gamma
%_{ij}-1 \right)^2}{2a}} + \right. \nonumber \\ && \left. + p_-(i,j)
%e^{ -\frac{\left(w_{ij}+a\Gamma _{ij}+1 \right)^2}{2a}} \right]
%\end{eqnarray}
\begin{eqnarray}
\label{binp} P\left( {\bf W}|D_M \right)=\frac{1}{(2\pi
a)^{N/2}}\prod _{ij}^N \left[ \hat{p}_{ij}^0  \ e^{
-\frac{\left(w_{ij}+a\Gamma _{ij} \right)^2}{2a}} + \right.
\nonumber \\ \left. + \hat{p}_{ij}^+ \
e^{-\frac{\left(w_{ij}+a\Gamma _{ij}-1 \right)^2}{2a}} +
\hat{p}_{ij}^- \ e^{ -\frac{\left(w_{ij}+a\Gamma _{ij}+1
\right)^2}{2a}} \right]
\end{eqnarray}
where $\hat{p}_{ij}^{\rm x}$ is the {\it a posteriori} probability
for gene $j$ regulate positively (${\rm x}=+$), negatively (${\rm
x}=-$) or to not regulate (${\rm x}=0$) gene $i$. These
probabilities are defined by $\hat{p}_{ij}^+=p_{ij}^{+}e^{ -\Gamma
_{ij}}/z_{ij}$, $\hat{p}_{ij}^-=p_{ij}^{-}e^{ \Gamma _{ij}}/z_{ij}$
and $\hat{p}_{ij}^0= p_{ij}^{0}/z_{ij}$, where $z_{ij}=1+ p_{ij}^+
\left( e^{-\Gamma _{ij}} -1\right) + p_{ij}^- \left(e^{\Gamma _{ij}}
-1\right)$ guarantee normalization. Furthermore, the relative
entropy of the {\it a posteriori} distribution Eq. \ref{ent} is
given by
\begin{equation}
H_r\left(D_M,P_0 \right)=-\sum_i^N I_g\left(i|D_M,P_0\right),
\end{equation}
where $I_g(i)$ is the information gain of gene $i$ with respect to
$P_0$ obtained from using the data $D_M$ which is defined by
%\begin{equation}
\begin{eqnarray}
I_g\left(i|D_M,P_0\right)&=& \sum _{j}^N\left[ \frac{a}{2} \Gamma
^2_{ij} - \ln
\left(z_{ij}\right) - \right. \nonumber \\ && \left.
\frac{1}{z_{ij}} \left(p_{ij}^+ \Gamma _{ij} e^{-\Gamma
_{ij}} - p_{ij}^ -\Gamma _{ij} e^{\Gamma _{ij}} \right)  \right]. \label{rent}
\end{eqnarray}
%\end{equation}

The multipliers $\Gamma _{ij}$ are obtained after solving the
equation
\begin{equation}
\langle w_{ij} \rangle = -a \Gamma _{ij} + z_{ij}^{-1} \left(
p_{ij}^+  e^{-\Gamma _{ij}} - p_{ij}^- e^{\Gamma _{ij}}\right).
\label{equ1}
\end{equation}
where $\langle w_{ij} \rangle$ are subject to the constraints
imposed by $D_M$. Our {\it central} idea is that of reinterpreting,
following %\cite{ld2},
the information in $D_M$ in a particular fashion,
\begin{equation}
\dot{{\bf X}}-{\bf B}= \langle {\bf W} \rangle \ {\bf X}.
\label{con}
\end{equation}
Thus, all of the possible networks that are consistent with Eq.
\ref{con}, can be written as
\begin{equation}\label{gsol}
\langle {\bf W} \rangle = \left( \dot{{\bf X}}-{\bf B} \right) \cdot
{\bf U} \cdot {\rm diag}(s_j^{-1}) \cdot {\bf V}^T + {\bf C}\cdot
{\bf V}^T
\end{equation}
${\bf C}=(c_{ij})$ is an $N \times N$ matrix, where $c_{ij}$ is zero
if $s_{j}\neq 0$ and is otherwise an arbitrary scalar coefficient.
${\bf U}$, ${\bf S}$ and ${\bf V}$ correspond to the singular value
decomposition of matrix ${\bf X}^T$, i.e. ${\bf X}^T={\bf U} \cdot
{\bf S} \cdot {\bf V} ^T$ where ${\bf U}$ is a unitary $M \times N$
matrix of left eigenvectors, ${\bf S}$ is diagonal $N\times N$
matrix containing the eigenvalues $\left\{ s_1,\ldots ,s_N
\right\}$, and ${\bf V}$ is a unitary $N\times N$ matrix of right
eigenvectors. Without loss of generality, let all non-zero elements
of $s_j$ be listed at the end and $s_j^{-1}$ in Eq. \ref{gsol} are
taken to be zero if $s_j=0$. The general solution \ref{gsol} can
be written as
\begin{equation}\label{l1}
\langle {\bf W} \rangle = {\bf W}_{L_2}+ {\bf C}\cdot {\bf V}^T
\end{equation}
where ${\bf W}_{L_2}$ is the particular solution with the smallest
$L_2$ norm. If $M<N$, many weights ${\bf W}$ are compatible with the
available information. The information contained in the data set
$D_M$ can be used in different ways. Each of these leads to a
different probability distribution which exhibits diverse
properties. In this sense, following the prescription $\langle {\bf
W} \rangle=0$ in Eq. \ref{l1}, the knowledge that gene regulatory
networks are sparse can be made use of. Thus, we have ${\bf C}\cdot
{\bf V}^T=-{\bf W}_{L_2}$, which is an overdetermined problem
\cite{yeung}. This particular solution will be denoted as ${\bf
W}_{L_1} $. Of course the ${\bf \Gamma}$ is obtained solving Eq.
\ref{equ1} using $\langle {\bf W} \rangle={\bf W}_{L_2} $ or
$\langle {\bf W} \rangle={\bf W}_{L_1} $. In the following sections
these alternatives will be considered independently. Notice that for
$M\geq N$, ${\bf W}_{L_2}={\bf W}_{L_1} $.

After determining the {\it a posteriori} distribution, the gene
interaction matrix $I$ must be selected. In order to do that, the
maximum likelihood criterion is taken into account, i.e. the
selection is accomplished choosing the highest {\it a posteriori}
probability from
$\{\hat{p}_{ij}^{0},\hat{p}_{ij}^{+},\hat{p}_{ij}^{-} \}$ for each
pair $i, j$. For example if $ \hat{p}_{ij}^+$ is greater than $
\hat{p}_{ij}^{0}$ and $\hat{p}_{ij}^{-}$, then $I_{ij}=1$ indicating
that gene $j$ activates the gene $i$.

In order to achieve the best model, the idea is to use the
information contained in $D_M$ and the knowledge that gene
regulatory networks are sparse. The formalism presented here offers
an alternative to the prescription which selects ${\bf W}_{L_1}$
from all possible solutions \ref{gsol}. This alternative consists
in setting $p_{ij}^+ = p_{ij}^- \ll p_{ij}^0$. In this way the
knowledge that gene regulatory network is sparse can be introduced
by assigning a much lower value to the {\it a priori} probabilities
of interaction than the {\it a priori} probabilities of absence of
interaction. Furthermore, as the inference processes occur row by
row, any other relevant {\it a priori} information about the gene in
consideration (such as known interactions, type of gene, etc.) could
be included in these probabilities. For example, if gene $k$ encode
a helix-turn-helix or a zinc finger protein, high probabilities can
be assigned for column $k$ ($p_{ik}^+$ and $p_{ik}^-$).

\section*{Results}

In order to systematically benchmark the inference performance of
this method, a linear data-generating model was used. The $M$ random
inputs (the columns of matrix ${\bf X}$) were generated in the range
[-1,1] and was computed ${\bf W} \cdot {\bf X}$ as the system
response, where ${\bf W}$ is the matrix to be reconstructed. Thus
pairs ${\bf {X,Y}}$ constitute the available information $D_M$. In
the simulation, it was observed that the mean performance depends on
size and the degree of connectivity and not on the network type. For
this reason, random sparse linear networks will be considered, where
each gene has $k$ entries in average. To build the connectivity
matrix ${\bf W}$, following procedure was used: for each matrix
element a random number $r$ between (0,1) was sorted, if $r<k/2N$ a
negative random value chosen from a uniform distribution was
assigned to the matrix element, if $r>1-k/2N$ the matrix element was
a positive random number, and otherwise the matrix element was zero.
The condition $k\ll N$ ensures sparseness.

By using singular value decomposition and interior point method for
$L_1$ regression ${\bf W}_{L_1}$ was computed. Subsequently the set
of uncoupled nonlinear equation \ref{equ1} was solved and the {\it
a posteriori} probability for each putative interaction was
evaluated. After this procedure the most likelihood ${\bf I}$ can be
selected. The performance of the inference procedure was measured by
the prediction error $\varepsilon = N^{-2} \sum _{ij}^N e_{ij}$,
where $e_{ij}$ is defined by
\begin{equation}
e_{ij}=\left\{ 0 \ \ {\rm if \ \ sign}(w_{ij})=I_{ij}
\atop 1  \ \ \ \ \ \ \ \ \ \ {\rm otherwirse} \right. .
\end{equation}

Figure 1 depicts the prediction error $\varepsilon$ as a function of
$\alpha$ defined as the ratio of number experiments and number of
genes, i.e. $\alpha =M/N$. These have been tested in three different
size networks with $k/N=0.05$, in which all {\it a priori}
probabilities are assumed to be equals (i.e.
$p_{ij}^+=p_{ij}^-=p_{ij}^0=1/3$) and $a=0.01$. For small values of
$M$ the method mistakenly infers a percentage of interaction which
depends on the network size $N$ and $k$. However, the prediction
error decays rapidly as $\alpha$ increases and the gene interaction
matrix is completely recovered with a $\alpha$ value that decreases
with the network size. This performance was obtained using ${\bf
W}_{L1}$ prescription. Similar simulations (data not shown)
performed with the ${\bf W}_{L2}$ prescription, reveal that in these
cases the prediction error $\varepsilon$ remains close to unit until
$\alpha =1$, where they decay abruptly. Dependence of performance on
the network topology has not been detected and similar results were
obtained for scale free networks which have more biological appeal
than the random networks used here. However, these simulations
present a greater error bar due to the fact that the network
building algorithm used here does not make networks with a uniform
node degree.

Many times, when dealing with an incomplete data set $M\ll N$, only
a percentage of the interactions is inferred correctly. If the
likelihood of the inferred paths cannot be assessed, this partial
reconstruction has small predictive value in real life. The
methodology proposed here can assess the likelihood of the predicted
interaction straightforwardly through the {\it a posteriori}
probability. In this sense, only those predicted interactions with
an {\it a posteriori} probability which is greater than some
significance level can be selected. To illustrate this issue, a
network with 60 genes with $k/N=0.05$ was simulated. The related
connectivity matrix ${\bf W}$ is represented in Fig. 2(left), row
$i$ corresponds to the genes that regulate the activity of gene $i$,
while column $j$ corresponds to the genes regulated by gene $j$. The
weight values $w_{ij}$ are depicted following a linear gray scale,
where white(black) corresponds to the maximum(minimum) values of
weights and the gray background represents the absence of
interaction. This network is random perturbed in 24 different
experiments ($\alpha= 0.4$). With this amount of data usually about
$\sim 99.5\%$ of the interactions are predicted correctly (see Fig.
1). Nevertheless, which interactions were inferred correctly and
which were inferred wrongly is unknown. By mean of the information
theory approach, the {\it a posteriori} probabilities were computed
and the inferred interaction matrix ${\bf I}$ and the associated
likelihood were derived. Fig. 2(right) represents the inferred
connectivity matrix ${\bf I}$, by assuming that all {\it a
priori} probabilities are equal (i.e. $p_{ij}^+=p_{ij}^-=p_{ij}^0=1/3$).
Red circles indicate wrong predictions ($1\%$ of the interactions),
while green circles indicate the interactions with a {\it a
posteriori} probability greater than 0.99. In this case there are
98(108) %(114)
interactions where the maximum {\it a posteriori} probability is
greater 0.99(0.95). %(0.90).
A more detailed study (data not show) revealed that interactions
related to higher weight values, are associated to high {\it a
posteriori} probability values. These results suggest that gene
networks can be partially recovered even with small amounts of data,
mainly for those genes that interact strongly.
%%%%%%

Unfortunately all measurements are subject to observational noise,
consequently it is important to asses to what extent the performance
of the inference procedure is affected by noise. To simulate this
condition in the numerical experiment, the available information
$D_M$ (both input and output) was corrupted by an additive Gaussian
noise with mean zero and standard deviation $\eta$. This inference
procedure was performed for networks with $N=60$, in the same
condition as for the previous assessment ($p_{ij}^+=p_{ij}^-=p_{ij}^0=1/3$ and $a=0.01$).
However, in this case the method based on the prescription of
sparseness assumed in $W_{L1}$ could not correctly recovery the gene
interaction matrix ${\bf I}$ when the noise level was $\eta=0.3$
(even for smaller $\eta$). Figure 3(top) indicates the prediction
error by using both ${\bf W}_{L1}$ and ${\bf W}_{L2}$ assuming that
the {\it a priori} probability for activation, repression or absence
of interaction are equal. This clearly shows that the prediction
power decreases as more data becomes available.

However, the network can be partially reconstructed by using an
alternative constraint of sparsity. This alternative consists in
introducing the knowledge of sparseness of the matrix through the
{\it a priori} probabilities. That is achieved by setting
$p_{ij}^{\pm}\approx 0$ in the inference procedure. Fig. 3(middle)
depicts the prediction error as a function of $\alpha$ when the {\it
a priori} probabilities were set to $p_{ij}^{\pm}=0.025$. The sum of
these probability values corresponds to the percentage of genes that
are regulated by one gene. With such {\it a priori} information, it
is possible to reconstruct almost the complete structure of the
network (around 95$\%$ of edges) using more experiments than the
number of genes, $\alpha\sim 2$. The mean node degree of the network
is generally not known in advance. Notwithstanding, the prediction
ability is robust for underestimations of the {\it a priori}
probabilities. Figure 3(bottom) depicts the prediction error as
function of $\alpha$ when the {\it a priori} probabilities were set
to $p_{ij}^{\pm}=0.01$. The result is almost the same as the
previous one. This implies that it is possible to partially recover
the interaction matrix even with noise data, by setting low values
for the {\it a priori} probabilities $p_{ij}^{\pm}$. In the last two
cases, the prediction performance obtained by the ${\bf W}_{L2}$
prescription is comparable with that obtained by ${\bf W}_{L1}$
using $p_{ij}^{\pm}\approx 0$ prescription, in contrast to the case
which deals with clean data. Furthermore, when data are corrupted by
noise, it was observed that prediction error has a peak around
$\alpha =1$, this peak arises because some singular values,
associated to the SVD, take small values as consequence of noise.

The partial recovery referenced above does not pursue recover a
closed subnetwork, which mainly infer strong interactions around the
whole network. However, in many cases this is crucial to recover the
complete subnetwork associated to a given gene or path of interest.
The inference approach and information gain tool presented in this
study, could be used to establish new relationships between genes
and to propose new experiments. By means of cycles of
experiments-datamining, the knowledge about the subnetwork can be
refined until its complete recovery, even in presence of
observational noise. For that purpose the following protocol could
be used: i) perform an initial perturbation where the gene of
interest is overexpressed, and obtain the genome expression profile;
ii) compute the information gain for each gene with this
experimental data; iii) select the genes for which the information
gain is greater than a given threshold; iv) iterate first two steps
perturbing each one of the genes which were selected in the third
step and which have still no been perturbed, until no new gene has
an information gain greater than the threshold. Figure 4A
illustrates the result of three of the experiments-datamining
cycles. Firstly, the gene which belongs to the subnetwork of
interest, gene g1, is initially overexpressed (level of 10.0 while
the other gene levels are random in the range [-0.5,0.5]), then the
input-output network is measured, this measurement is subject to
observational noise with $\eta=0.30$. The information gain of this
experiment is computed for each gene using $p_{ij}^+=p_{ij}^-=0.01$
as an {\it a priori} probability. Subsequently, those genes with
$I_g$ greater than 1.0 are selected. $I_g$ suggests that gene g6 is
regulated by g1. Repeating the above step with gene g6, the results
indicate that genes g2, g3 are regulated by g6. The above step is
repeated with gene g2 and subsequent genes with high information
gain values in ensuing experiments, until no new gene with an
information gain greater than threshold appears. Fig. 4B illustrates
a list of experiments where the first column corresponds to the gene
that was perturbed in the experiment, and the second column
corresponds to the genes which appear to be regulated by the
perturbed gene. In the last two experiments no new regulated genes
appeared (which were not indicated in the first column list). The
above analysis provides a causal link between two genes, but it does
not indicate if the regulation is positive or negative. In order to
extract this information, the inference analysis was performed using
the ten "overexpression experiments" pooled in $D_M$ ($M=10$). When
the inference procedure was applied with this data, 19 out of 24
interactions in the subnetwork were inferred correctly, 10 of them
with an {\it posteriori probability} greater than 0.99. However, the
{\it a priori} probabilities provided by the information contained
in list of Fig. 4B are included, setting $p_{ij}^+=p_{ij}^-=0.5$ (or
1/3) for all the pairs $i,j$ indicated in the list, and
$p_{ij}^+=p_{ij}^-=0.01$ otherwise, 23 out of 24 interactions in the
subnetwork are inferred, 19 of them with {\it a posteriori}
probability greater than 0.99, Fig. 4C. The performance above
obtained does not differ if the inference procedure is implemented
using ${\bf W}_{L2}$ or ${\bf W}_{L1}$ prescription, of course ${\bf
W}_{L2}$ is computationally cheaper than ${\bf W}_{L1}$ which
requires linear programming optimization.

The above example about subnetwork inference suggests that this
novel scheme can be re-used regarding further subnetworks until the
whole network is recovered with $M\simeq N$ experiments.

\section*{Discussion and Conclusions}

A novel approach for regulatory network inference is presented in
this study. Differently to other methods, this approach pursues to
infer the type of interaction rather than a weight which
characterizes the interaction quantitatively. Three main features of
the proposed method are pointed out. First, it allows to introduce
global {\it a priori} information about the network, as sparseness,
and other gene dependent available information, as illustrated in
the last example Fig. 4C. Second, the information theory formalism
provides a way to quantify the likelihood of the inferred paths, by
using the {\it a posteriori} probabilities computed with the method.
Last, but not least, information theory formalism also quantifies
the information gained with the set of data to be used in the
inference procedure.

Furthermore, the IT approach seems to offer promising perspective as
a network inference protocol; the methodology presented here
introduces an information gain measure as a bonus. The way in which
this quantity could be a useful tool to identify the downstream
regulated genes in overexpression experiments is illustrated in this
study. This feature allows a datamining-assisted way of uncovering
the whole network with a number of experiments equal to the number
of genes, even when dealing with a high level of observational
noise. This IT approach enables the effective use of all the
available information, in which each experiment is used as an
individual constraint. Thus, the ensuing observation level becomes
much richer than the standard one, where all data define a fitness
function to be optimized. Efficient management leads to more
realistic results in inference.

The learning protocol presented here constitutes an additional
inference technique, which should be of interest not only for basic
research but also as an application to many interesting real world
problems without paying an excessive computational cost.

\section*{Acknowledgments}
The author thanks Christina McCarthy for critical review of the manuscript.
L.D. is researcher of CONICET (Argentina).

\newpage

\begin{figure}
\centerline{\includegraphics[width=.5\textwidth]{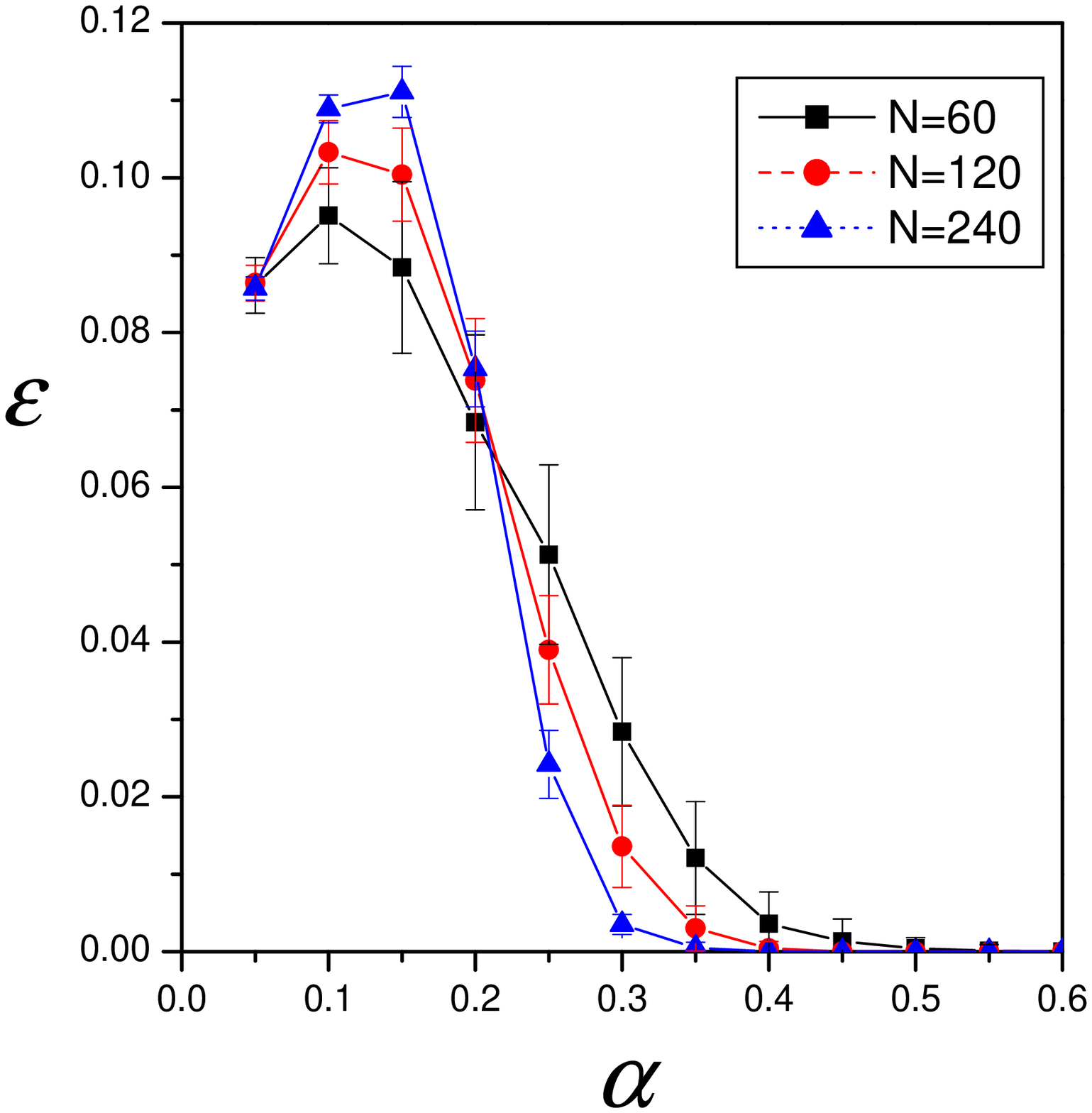}}
\caption{(Color online) Performance. Prediction error $\varepsilon$ as a function
of the ratio $\alpha=M/N$ for gene networks with 60 genes (squares),
120 genes (circles) and 240 genes (triangles), averaged over 50
networks. In all cases the performances were obtained using $W_{L1}$
prescription, equal {\it a priori} probabilities (i.e.
$p_{ij}^+=p_{ij}^-=p_{ij}^0=1/3$ for all $i$ and $j$), $k/N=0.05$
and $a=0.01$.}
\end{figure}
\newpage

\begin{figure*}
\centerline{\includegraphics[width=1.0\textwidth]{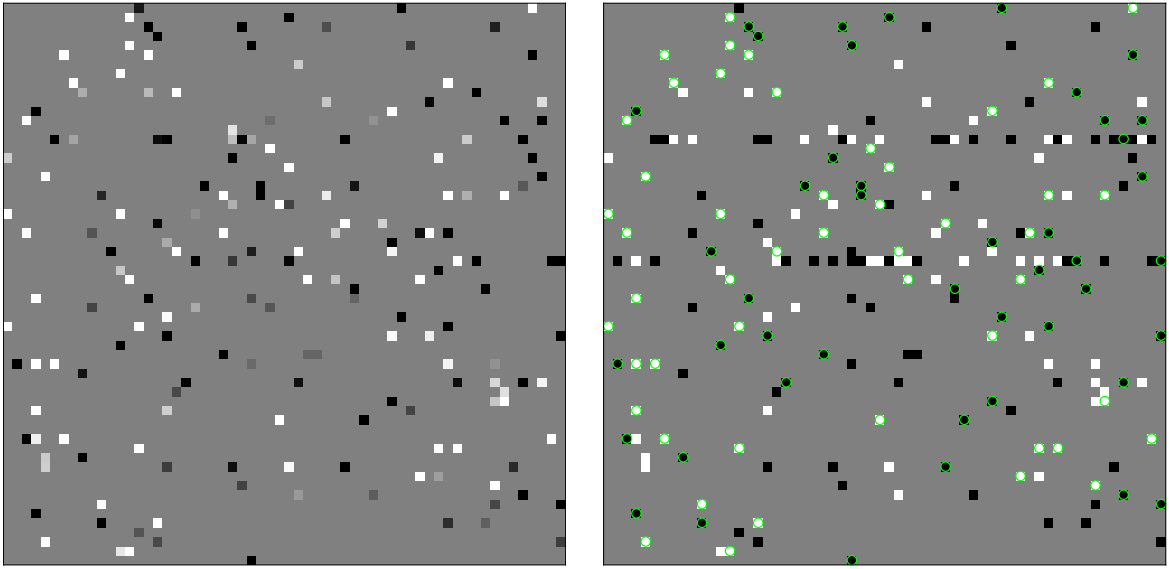}}
\caption{(Color online) Likelihood assessment. Left: connectivity matrix ${\bf W}$
representation related to a random network of 60 genes with
$k/N=0.05$. Rows correspond to regulated genes, while columns
correspond to the genes acting as regulators. The interaction
weights $w_{ij}$ are represented following a linear gray scale,
where white corresponds to $w_{ij}=2$, while black to $w_{ij}=-2$.
Gray background represents the absence of interaction, i.e.
$w_{ij}=0$. Right: gene interaction matrix ${\bf I}$ inferred after
24 random perturbation experiments, using $W_{L1}$ prescription,
$a=0.01$ and $p_{ij}^+=p_{ij}^-=p_{ij}^0=1/3$.
Circles (green in the online figure) indicate the 94 interactions with an {\it a
posteriori} probability greater than 0.99. %Red circles indicate
Wrong predictions (35, $\sim 1\%$ of the putative interactions, which in this case)
correspond to the regulatory inputs of two genes.}
\end{figure*}
\newpage

\begin{figure}%
\centerline{\includegraphics[width=0.50\textwidth]{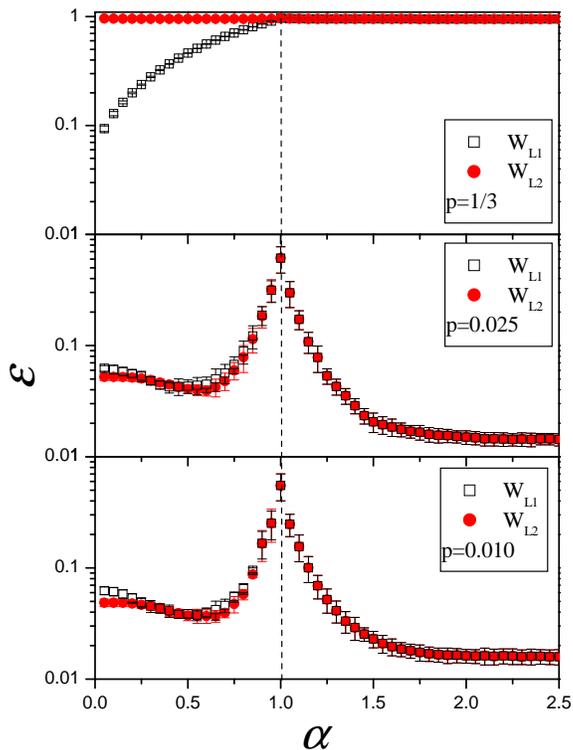}}
\caption{(Color online) Inferring with noisy data. Prediction error $\varepsilon$
as a function of the ratio $\alpha$ for gene networks with 60 genes
with $k/N=0.05$. Both input and output data are subject to
observational noise of $\eta=0.30$. The performance was obtained
using both $W_{L1}$ (open square) and $W_{L2}$ (filled
circle) prescriptions and $a=0.01$. Top: the {\it a priori}
probabilities are equal, i.e. $p_{ij}^+=p_{ij}^-=p_{ij}^0=1/3$ for
all $i$ and $j$. Medium: the {\it a priori} probabilities are set to
be $p_{ij}^+=p_{ij}^-=0.025$ and $p_{ij}^0=0.95$ for all $i$ and
$j$. Bottom: the {\it a priori} probabilities are set to be
$p_{ij}^+=p_{ij}^-=0.01$ and $p_{ij}^0=0.98$ for all $i$ and $j$.}
\end{figure}
\newpage

\begin{figure*}
\centerline{\includegraphics[width=1.0\textwidth]{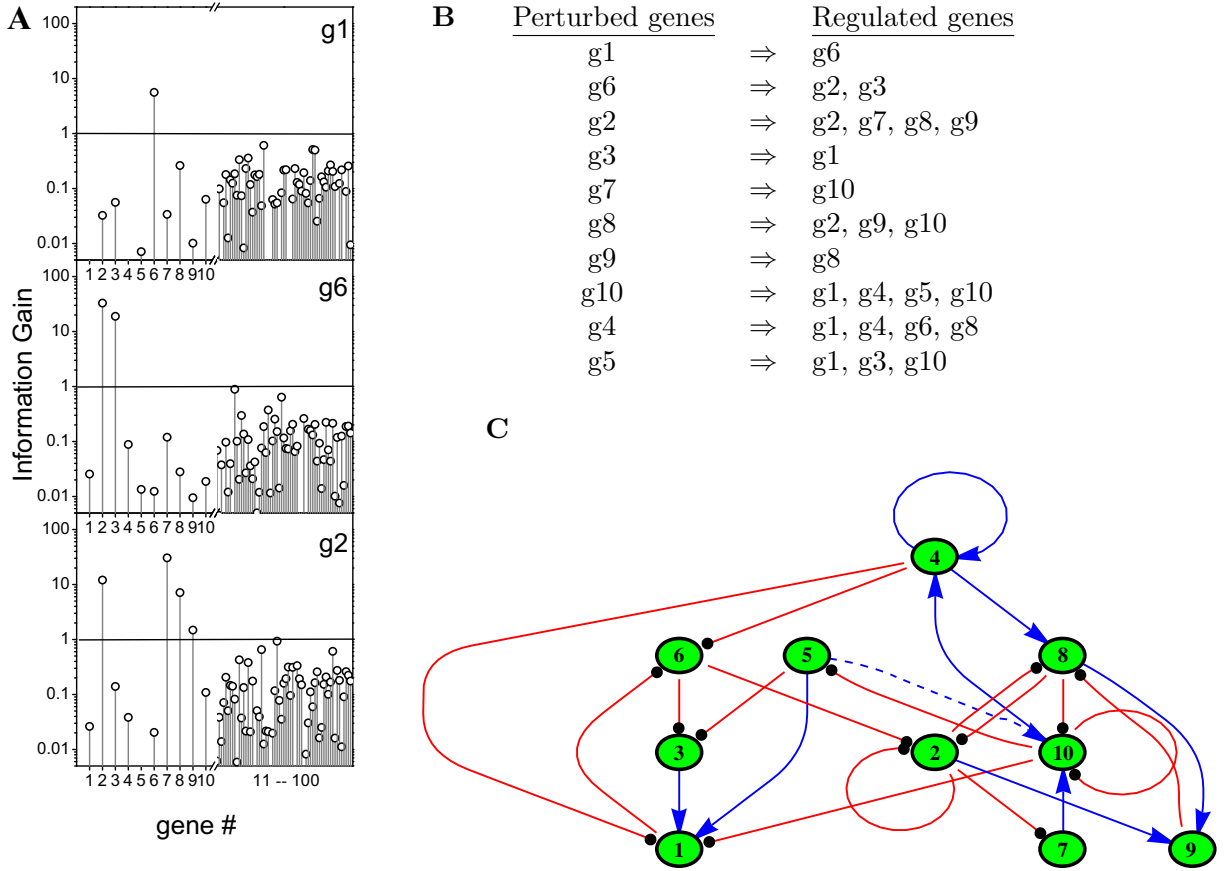}}
\caption{(Color online) Subnetwork identification. A: Information gain $I_g$
obtained for three ``overexpression experiments". Firstly, the gene
which belongs to the subnetwork of interest, gene g1, is initially
overexpressed, then the input-output network is measured, this
measurement is subject to observational noise of $\eta=0.30$. The
information gain of this experiment is computed for each gene and
the genes with $I_g$ greater than a given threshold, are selected.
$I_g$ suggests that gene g6 is regulated by g1. Repeating the above
step with gene g6, it appears that genes g2, g3 are regulated by g6.
The above step is repeated with gene g2 and subsequent genes with
high information gain values in subsequent experiments. B: List of
experiments, the first column corresponds to the gene which was
overexpressed in each experiment, the second column corresponds to
the genes which appear to be regulated by the overexpressed gene. C:
Subnetwork inferred 23 out of 24 interactions correctly (solid
edges) by this inference procedure using $W_{L2}$ prescription and
the above ten "overexpression experiments" together. The information
contained in list B was included as {\it a priori} probabilities,
i.e., they were set $p_{ij}^+=p_{ij}^-=0.5$ and $p_{ij}^0=0.0$ for
all $i,j$ pairs indicated in the list, and
 $p_{ij}^+=p_{ij}^-=0.01$ otherwise.}
\end{figure*}

\end{document}